\documentclass[
twocolumn,
aps,
pre,
preprintnumbers,
amsmath,
amssymb
]{revtex4}
\usepackage{graphicx}	
\usepackage{dcolumn}	
\usepackage{bm}			

\begin{document}
\title{Can Local Stress Enhancement Induce Stability in Fracture Processes? Part I: Apparent Stability}

\author{Jonas T. Kjellstadli}
\email{jonas.kjellstadli@outlook.com}
\author{Eivind Bering}
\email{eivind.bering@ntnu.no}
\author{Martin Hendrick}
\email{martin.hendrick@ntnu.no} 
\author{Srutarshi Pradhan}
\email{srutarshi.pradhan@ntnu.no}
\author{Alex Hansen}
\email{alex.hansen@ntnu.no}
\affiliation{PoreLab, Department of Physics, NTNU -- Norwegian University of Science and Technology, Trondheim, Norway}

\date{\today}

\begin{abstract}
By comparing the evolution of the local and equal load sharing fiber bundle models, we point out the paradoxical result that stresses seem to make the local load sharing model stable when the equal load sharing model is not. We explain this behavior by demonstrating that it is only an \textit{apparent stability} in the local load sharing model, which originates from a statistical effect due to sample averaging. Even though we use the fiber bundle model to demonstrate the apparent stability, we argue that it is a more general feature of fracture processes.
\end{abstract}
\maketitle

\section{Introduction}
\label{intro}

The stability of materials against fracture is essential for our civilization. We need to be able to trust that buildings, bridges, airplanes, ships, etc.\ do not collapse. To prevent the collapse of structures, one needs to understand the processes that constitute fracture. Fracture has been studied by the engineering and materials science communities for a very long time \cite{l93}. Only over the last thirty years, it has also entered physics \cite{hr90}. Within the physics approach to fracture, there has been an emphasis on the role of disorder and fluctuations \cite{rh90,cb97}.

We may summarize the physics of fracture in a heterogeneous brittle materials as follows: The material heterogeneity implies that both the local strength of the material and the stress field it is experiencing are themselves heterogeneous. Fractures may occur and develop as a result of either the material being locally weak or locally under high stress. Applying a sufficiently large load to a material, the fracture process will start by the material failing where it is weakest. The ensuing microcracks will induce high stresses at the crack tips. If these are sufficiently high, the microcracks will grow. Hence, a competition between stress enhancement due to developing microcracks and local material weakness breaks out \cite{rh90,cb97,bb11}. At some point, the stress intensity at the crack tips has become so large that the local material weakness is no longer able to compete and catastrophic failure sets in: a macroscopic crack develops. 

Essential in this summary is the opposite roles played by heterogeneity and stress enhancement: the heterogeneity stabilizes the fracture process whereas the stress enhancement destabilizes it. In this paper we demonstrate that stress enhancement may seemingly have the \textit{opposite} effect, i.e., it stabilizes the fracture process. This is a situation which essentially turns upside down common wisdom within the physics community on how fracture processes proceed.

It turns out, however, that this paradoxical behavior is an apparent effect caused by the fluctuations that occur during the fracture process. We use the fiber bundle model \cite{d45,phc10,hhp15} to demonstrate the \textit{apparent stability} and its explanation. We consider two variants of the model: the \textit{equal load sharing} (ELS) model \cite{p26} where there is local heterogeneity but no local stress enhancement, and the \textit{local load sharing} (LLS) model \cite{hp78} where there is a competition between local stress enhancement and local heterogeneity. Even though we use the fiber bundle model as a tool to demonstrate the apparent stability, we argue that the effect is more general. The lesson to be learned is the following: even though the average stress vs.\ strain curve may have a positive slope, seemingly indicating stability, the positive slope is not necessarily caused by stability, but by the evolution of the fluctuations biasing the average in a way that makes the slope positive.

However, there also exists a real effect where the local stress enhancement of the LLS model can make it more stable than the ELS model. This shielding effect --- its origins and consequences --- is the subject of Part II \cite{kbph19}.

There are two main sources of fluctuations in dynamical systems such as materials failing under stress \cite{fw12,a15}: one comes from statistical fluctuations of the probability distributions that define intrinsic properties of the system elements. Another type of fluctuations arises as a result of the system dynamics depending on the 
spatial structures. The first type of fluctuation has a direct relation with the system size and it normally disappears as the system size diverges due to self averaging. One can minimize the effect of these fluctuations either by making the system size larger or by increasing the number of samples. On the other hand, the dynamics-dependent fluctuations do not disappear with increasing size. It is therefore crucial to know the nature of this second type of fluctuations and its role during the entire evolution dynamics. It is this second type of fluctuations that is the cause of the apparent stability.

\section{The Fiber Bundle Model}
\label{fbm}

A fiber bundle consists of $N$ fibers placed between two clamps. The fibers act as Hookean springs with identical spring constants $\kappa$ up to an extension threshold $t_{i}$, individual for each fiber $i$, where they fail and cannot carry a load any more. Hence the connection between the extension $x$ of a fiber $i$ and the force $f_{i}$ it carries is
\begin{equation}
\begin{aligned}
f_{i} = 
\begin{cases}
\kappa x & \text{if } x < t_{i}, \\
0 & \text{if } x \geq t_{i}.
\end{cases}
\end{aligned}
\end{equation}
The thresholds $t_{i}$ are drawn from a probability density $p(t)$, with corresponding cumulative probability $P(t) = \int_{0}^{t} p(u) \text{d}u$.

\subsection{Equal Load Sharing}
\label{els}

In the ELS model an externally applied force $F$ is distributed equally on all the intact fibers. This means that fibers fail in order of increasing thresholds as the force $F$ increases. The force per fiber $\sigma = F/N$ required to give the bundle an extension $x$ is on average \cite{hhp15}
\begin{equation}
\sigma(x) = \kappa \left(1-P(x)\right)x.
\label{eq:els_load_curve}
\end{equation}
Equivalently,
\begin{equation}
\sigma(k) = \kappa \left(1-\frac{k}{N}\right) P^{-1}\left(\frac{k}{N}\right),
\label{eq:els_force_dist}
\end{equation}
since $P(x)$ is the fraction of broken fibers $k/N$ --- also called the damage $d$ --- at extension $x$ \cite{g04}. The fluctuations around this average are of the first type, and disappear as $N^{-1/2}$ when $N \to \infty$ \cite{hhp15}.

The \textit{load curve} is the smallest force per fiber $\sigma$ required to break the next fiber. Hence, we plot either this minimum $\sigma$ as a function of the extension $x$ or the fraction of broken fibers $k/N$, see Fig.\ \ref{fig1}. When plotted against the extension $x$, the load curve is the stress-strain curve. Equations (\ref{eq:els_load_curve}) and (\ref{eq:els_force_dist}) give the average load curve for ELS. We will use the terminology that a fiber bundle is \textit{locally stable} if the load $\sigma$ must be increased to continue breaking more fibers, i.e., if the load curve is increasing. From equation (\ref{eq:els_load_curve}) we determine the critical extension $x_c$ at which the ELS model becomes unstable by setting $\text{d}\sigma/\text{d}x|_{x_c}=0$. For a general Weibull threshold distribution
\begin{equation}
P(t) = 1 - \exp(-t^{\beta}+t_{0}^{\beta})
\label{eq:Weibull}
\end{equation}
with shape parameter $\beta$ and lower cut-off $t_{0}$ ($t \geq t_{0}$), this gives $x_{c} = \beta^{-1/\beta}$. This means that the ELS model is unstable from the beginning of the failure process when $t_{0} \ge \beta^{-1/\beta}$. 

\subsection{Local Load Sharing}
\label{lls}

In the LLS model, the loads originally carried by broken fibers are carried by their nearest intact neighbors only. Hence there is a spatially dependent stress field. A \textit{hole} is a cluster (in the percolation sense) of $h$ failed fibers joined through nearest neighbor connections. The \textit{perimeter} of a hole is the set of $p$ intact fibers that are nearest neighbors of the hole. With these definitions the force acting on an intact fiber $i$ with the LLS model is given by
\begin{equation}
f_{i} = \sigma \left(1+\sum_{j}\frac{h_j}{p_j}\right),
\label{eq:lls_force_dist}
\end{equation}
where $j$ runs over the set of holes that neighbor fiber $i$. The first term is the force originally applied to every fiber, while the second is the redistribution of forces due to failed fibers. Equation (\ref{eq:lls_force_dist}) is completely general, and can be used for any lattice and dimensionality, or even for random graphs.

To determine which fiber breaks next under an external load we define the \textit{effective threshold} $t_{\text{eff,}i}$ of fiber $i$ as
\begin{equation}
t_{\text{eff,}i} = \frac{t_{i}}{1+\sum_{j}\frac{h_{j}}{p_{j}}}.
\label{eq:effective_threshold}
\end{equation}
The breaking criterion of fiber $i$ is then $\sigma = \kappa t_{\text{eff,}i}$, and the fiber with the smallest effective threshold will fail under the smallest applied load $\sigma$.

\section{Determining Stability}
\label{stability}

\begin{figure}
\begin{center}
\includegraphics[width=\columnwidth]{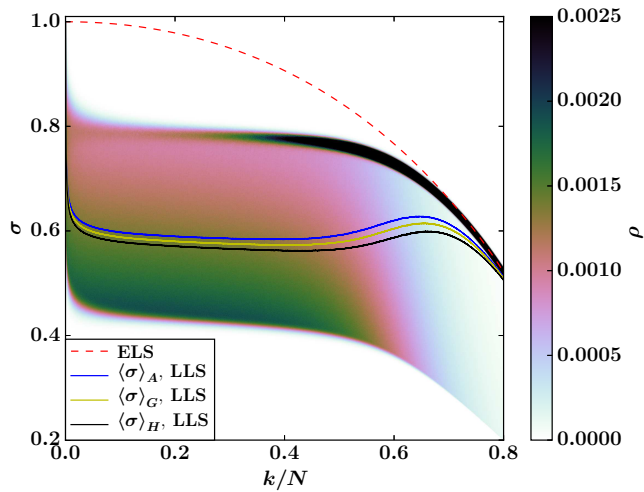}
\caption{Load curves for the ELS and LLS models with a Weibull threshold distribution $P(t) = 1 - \exp \left( -t + 1 \right)$, equation (\ref{eq:Weibull}) with $\beta = t_{0} = 1$. The ELS curve is equation (\ref{eq:els_force_dist}), whereas the LLS curves are sample averages --- arithmetic mean $\left< \sigma \right>_{A}$, geometric mean $\left< \sigma \right>_{G}$, and harmonic mean $\left< \sigma \right>_{H}$ --- from simulations on a square lattice ($N = 128^{2}$). The background is a color map that shows the density $\rho$ of single sample LLS load curves for the $1.5 \times 10^{5}$ samples that the averages are based on. The color bar is capped at $\rho = 0.0025$ to highlight the fluctuations with the smallest values of $\sigma$.}
\label{fig1}
\end{center}
\end{figure}

The LLS model contains stress enhancement in that fibers belonging to the perimeters of holes carry more load than corresponding fibers in the ELS model. Therefore the results in Fig.\ \ref{fig1} --- where we show the load curves ($\sigma$ vs.\ $k/N$) of the ELS and LLS model based on the the Weibull threshold distribution from equation (\ref{eq:Weibull}) with $\beta = t_{0} = 1$ --- are surprising. The ELS load curve is unstable for all values of $k/N$ (as indicated by the negative slope) because $t_{0} = x_{c}$, but there is a region for which the sample averaged LLS load curve has a positive slope, which seems to indicate local stability. This was first pointed out by Sinha et al.\ \cite{skh15}.

Our explanation of this paradoxical behavior lies in the difference between single samples and sample averages. Stability is a property of individual samples, not the average behavior. In the ELS model there is no difference between the two, since fluctuations around the sample averaged load curve are of the first type and disappear as $N \rightarrow \infty$. But the LLS model has fluctuations of the second type --- they persist in the limit of infinitely large systems --- and we must therefore study individual samples to determine when systems are stable.

We argue that in the LLS model, the stability of single samples, both global and local, is determined by the upper bounding curve of the force fluctuations. We show in Fig.\ \ref{fig1} how the density $\rho$ of these fluctuations are distributed around the averaged load curve for the LLS model. In any finite, but small, damage interval $[k/N,k/N+\Delta]$ there will be at least one strong fiber that requires a load $\sigma$ close to the upper bounding curve to break. For a system to be locally stable, consecutive intervals must require higher loads to break, i.e., the bounding curve of the fluctuations must increase. We see in Fig.\ \ref{fig1} that it does not. Hence, there is no local stability for the LLS model either, as expected.

Are other averages than the arithmetic mean more representative of individual samples? In the field of Anderson localization \cite{a58}, the average conductance differs vastly between different averaging procedures \cite{smo01}, and the arithmetic mean is not representative of typical single samples. We therefore show both the arithmetic, geometric, and harmonic mean of the LLS model in Fig.\ \ref{fig1}. The three means all give qualitatively similar behavior, and all of them fail to represent the behavior of single samples. Hence we will from now on use $\left< \sigma \right>$ for the arithmetic mean $\left< \sigma \right>_{A}$, which is a suitable representative for the three means when computing sample averages.

\section{Apparent Stability}
\label{apparent_stability}

The \textit{apparent} local stability in Fig.\ \ref{fig1} is caused by the sharp decline in fluctuations smaller than the average in a damage region around the site percolation threshold $p_{c} \approx 0.59$ \cite{sa92} of the square lattice.

\begin{figure}
\begin{center}
\includegraphics[width=\columnwidth]{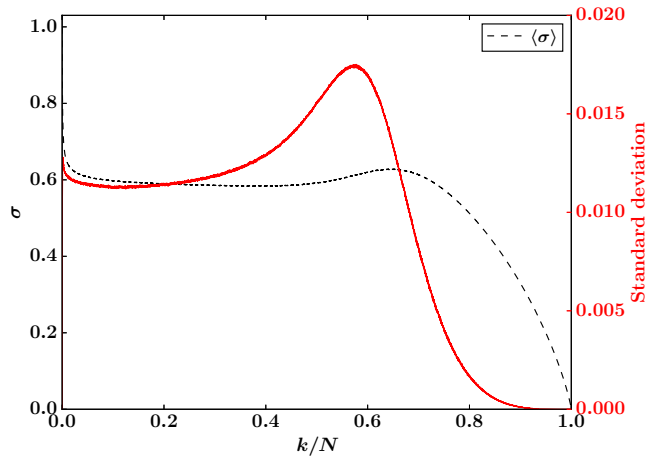}
\caption{Sample averaged load curve for the LLS model (black, left axis) and corresponding standard deviation (red, right axis) for a Weibull threshold distribution with $\beta = t_{0} = 1$. Results are from simulations on a square lattice ($N = 128^{2}$) with $1.5 \times 10^{5}$ samples.}
\label{fig2}
\end{center}
\end{figure}

The fluctuations are initially heavily biased with a large concentration below the average. Around the percolation threshold, the bias in the fluctuations begins to shift rapidly from small values of $\sigma$ to the upper bounding curve, and this shift is enough to make the average load curve increase even though the upper bounding curve is decreasing. This is supported by Fig.\ \ref{fig2}, which shows the averaged LLS load curve and its standard deviation.

We now consider a uniform threshold distribution on $[t_{0},1)$:
\begin{equation}
P(t) = \frac{t-t_{0}}{1-t_{0}},
\label{eq:uniform}
\end{equation}
which gives a critical extension $x_{c} = 1/2$. The ELS model with this distribution is hence unstable from the beginning of the breaking process if $t_{0} \geq 1/2$. We choose $t_{0} = 1/2$ for a comparable situation to the Weibull distribution studied earlier.

\begin{figure}
\begin{center}
\includegraphics[width=\columnwidth]{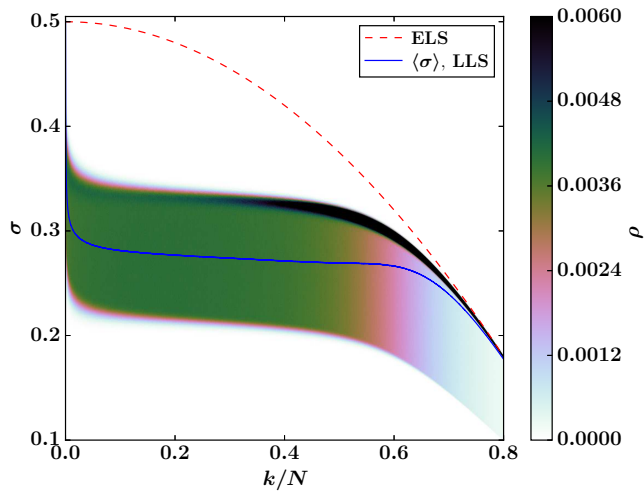}
\caption{Load curves for the ELS and LLS models with a uniform threshold distribution on $[0.5,1)$, equation (\ref{eq:uniform}) with $t_{0} = 1/2$. The ELS curve is equation (\ref{eq:els_force_dist}), whereas the LLS curve is a sample average from simulations on a square lattice ($N = 128^{2}$). The background is a color map that shows the density $\rho$ of single sample LLS load curves for the $1.5 \times 10^{5}$ samples that the average is based on. The color bar is capped at $\rho = 0.006$ to highlight the fluctuations with the smallest values of $\sigma$.}
\label{fig3}
\end{center}
\end{figure}

The averaged load curves for LLS and ELS with the threshold distribution from equation (\ref{eq:uniform}) are shown in Fig.\ \ref{fig3}, together with the density of fluctuations around the LLS load curve. The upper bounding curve of the LLS force fluctuations decreases for all damages $k/N$, indicating that the system is unstable throughout the breaking process also for this uniform threshold distribution.

\begin{figure}
\begin{center}
\includegraphics[width=\columnwidth]{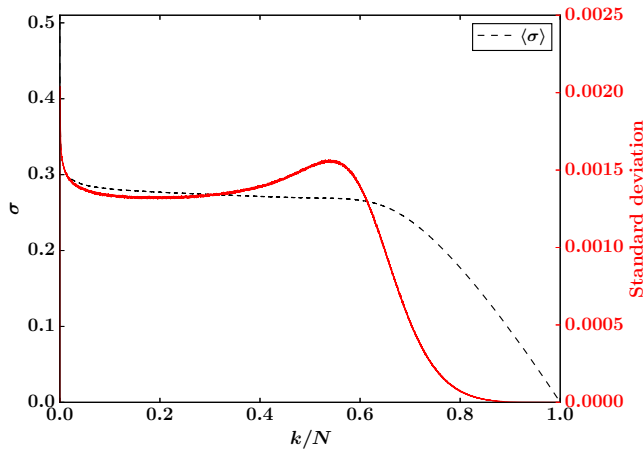}
\caption{Sample averaged load curve for the LLS model (black, left axis) and corresponding standard deviation (red, right axis) for a uniform threshold distribution on $[0.5,1)$. Results are from simulations on a square lattice ($N = 128^{2}$) with $1.5 \times 10^{5}$ samples.}
\label{fig4}
\end{center}
\end{figure}

In Fig.\ \ref{fig3} there is a region around the percolation threshold where the fluctuations shift --- similarly to how they change in Fig.\ \ref{fig1} for the Weibull threshold distribution --- corroborated by the standard deviation in Fig.\ \ref{fig4}. In this case the fluctuations are not very biased to begin with, but distributed almost uniformly around the average. This --- and the fact that the fluctuations span a smaller range of forces $\sigma$, as demonstrated by a standard deviation an order of magnitude smaller in Fig.\ \ref{fig4} than in Fig.\ \ref{fig2} --- makes the shift of the fluctuations smaller than for the Weibull threshold distribution, and it is not enough to make the averaged load curve increase as a function of damage.

The averaged LLS load curve does not show any apparent stability for the uniform distribution, but the underlying effect --- that the distribution of the force fluctuations changes rapidly in a region around the percolation threshold --- that causes the apparent stability for the Weibull distribution is still present, as shown by the color map in Fig.\ \ref{fig3}.

\begin{figure*}
\begin{center}
\includegraphics[width=0.3\textwidth]{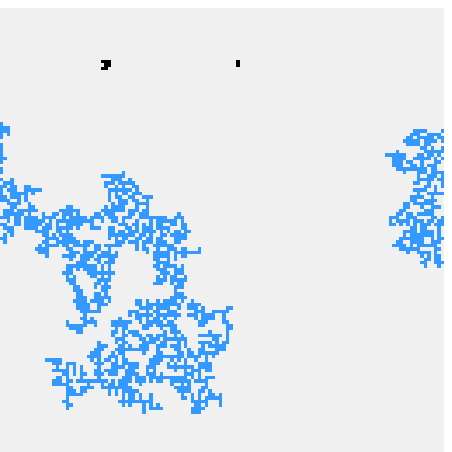}
\includegraphics[width=0.3\textwidth]{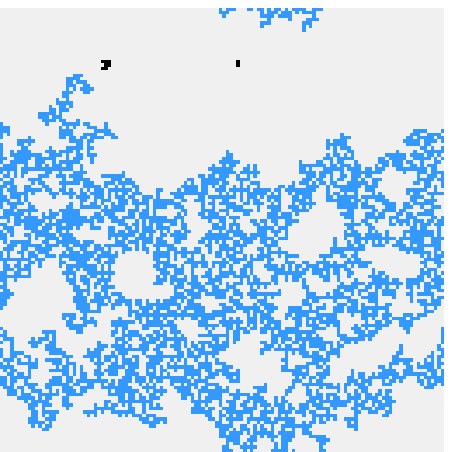}
\includegraphics[width=0.3\textwidth]{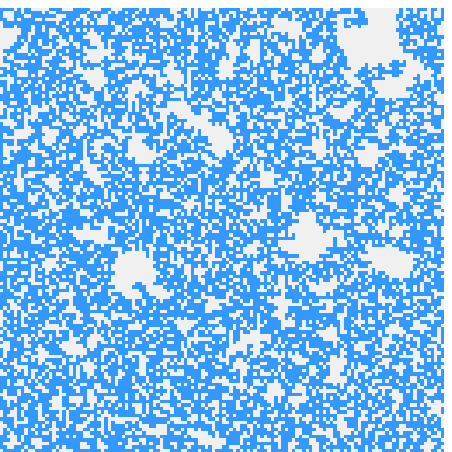}
\caption{Hole structure of a square lattice ($N = 128^{2}$) LLS fiber bundle with the Weibull threshold distribution $P(x) = 1 - \exp \left( -x + 1 \right)$ at three different damages: $k/N = 0.1$ (left), $k/N = 0.3$ (middle) and $k/N = 0.59$ (right). Intact fibers are light gray, the largest hole is blue, and other broken fibers are black. From early on in the breaking process a single hole is growing continually.}
\label{fig5}
\end{center}
\end{figure*}

\begin{figure*}
\begin{center}
\includegraphics[width=0.3\textwidth]{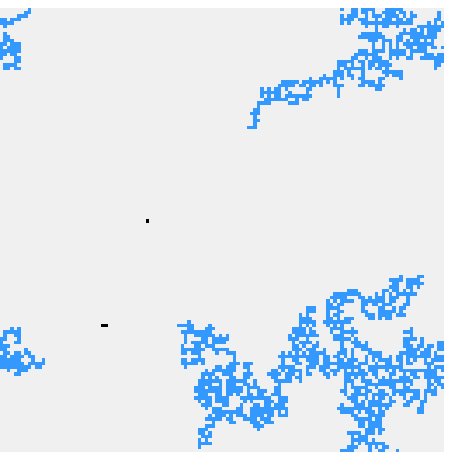}
\includegraphics[width=0.3\textwidth]{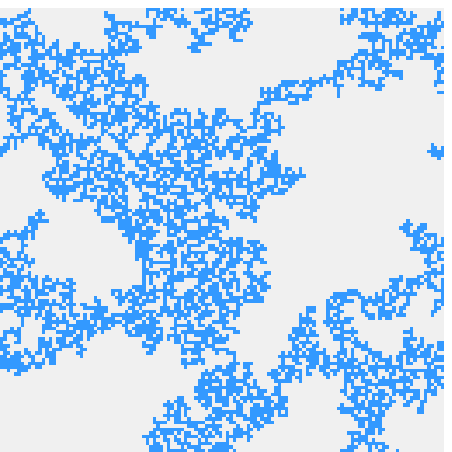}
\includegraphics[width=0.3\textwidth]{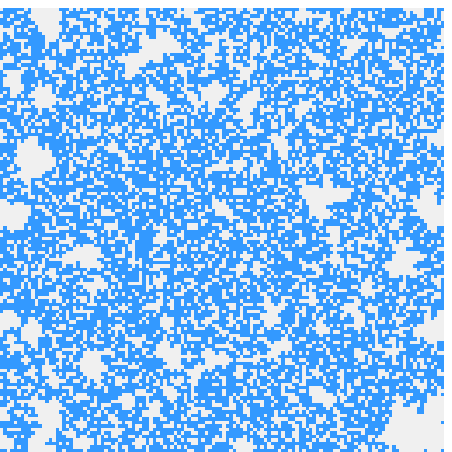}
\caption{Hole structure of a square lattice ($N = 128^{2}$) LLS fiber bundle with a uniform threshold distribution on $[0.5,1)$ at three different damages: $k/N = 0.1$ (left), $k/N = 0.3$ (middle) and $k/N = 0.59$ (right). Intact fibers are light gray, the largest hole is blue, and other broken fibers are black. From early on in the breaking process a single hole is growing continually.}
\label{fig6}
\end{center}
\end{figure*}

The changing fluctuations around the percolation threshold can be understood by examining the hole structure of the LLS fiber bundle as the damage increases. When only a few fibers have broken the breaking process localizes around a single hole, which starts expanding and keeps growing until the entire fiber bundle has broken \cite{b16}. This growth process is illustrated in Fig.\ \ref{fig5} for the Weibull threshold distribution and Fig.\ \ref{fig6} for the uniform threshold distribution.

Fibers that break after the localization sets in are in the perimeter of the growing hole. Since there are almost no other holes, (nearly all) perimeter fibers get the same contribution to effective threshold from the hole structure (Eq.\ (\ref{eq:effective_threshold})), and therefore the fiber that breaks next is the perimeter fiber with the smallest threshold. This results in a breaking process that is similar to invasion percolation where the weakest neighboring fiber is ``invaded'' by the hole every time a fiber breaks.

Hole and perimeter sizes of the growing hole are similar in different samples. Hence the force fluctuations in Figs.\ \ref{fig1} and \ref{fig3} mainly represent the distribution (over samples) of the smallest threshold in the perimeter of the hole. It follows that the lower end of the fluctuations are due to the hole encountering new fibers with small thresholds as it expands. These fibers break quickly --- they are likely to have the smallest threshold among the perimeter fibers --- while the stronger fibers in the perimeter survive.


The lower end of the fluctuations disappear rapidly around the percolation threshold because the growing hole has permeated most of the lattice, and therefore has few new areas to expand into, as shown in Figs.\ \ref{fig5} and \ref{fig6}. As a result, there are few new neighborhoods to expand into to find new neighbors with small thresholds. This mechanism radically changes the distribution of force fluctuations, so that the sample averaged load curve increases in Fig.\ \ref{fig1} even though individual samples are all locally unstable.

\subsection{The Effect of the Lattice}

\begin{figure}
\begin{center}
\includegraphics[width=\columnwidth]{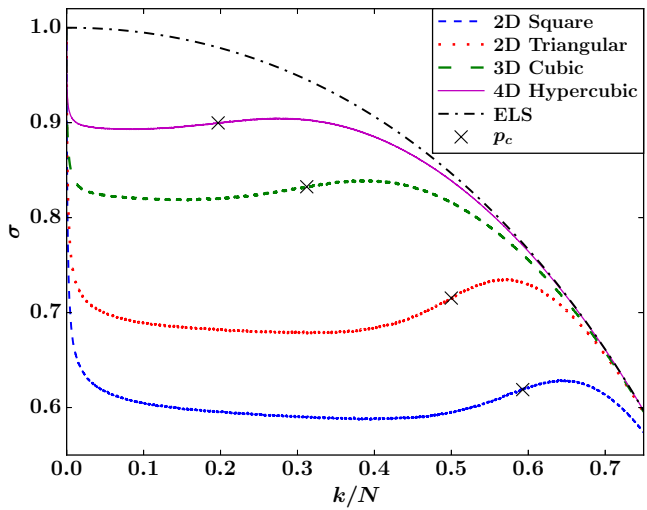}
\caption{Sample averaged LLS load curves on lattices in two to four dimensions with corresponding site percolation thresholds $p_{c}$ marked. The threshold distribution is $P(x) = 1 - \exp \left( 1 - x \right)$, equation (\ref{eq:Weibull}) with $\beta = t_{0} = 1$.}
\label{fig7}
\end{center}
\end{figure}

The above reasoning does not hinge on the lattice being square, and should be valid for any lattice. We therefore expect that the same Weibull threshold distribution will give similar results for the LLS model on other lattices: the sample averaged load curve should increase around the site percolation threshold due to the shift in bias as the lower end of the fluctuations disappear. We show averaged LLS load curves for four lattices in 2D, 3D and 4D for the Weibull threshold distribution in Fig. \ref{fig7}. The figure shows positive slopes of the load curves for all four lattices in a region around the corresponding percolation threshold, in accordance with the above argument. In all of these cases, individual samples are locally unstable, showing that the sample averaged load curve cannot be trusted as an indicator of local stability.

\subsection{Apparent Stability in Globally Stable Systems}

From the examples presented so far, it could be argued that the effect we describe is less relevant because it occurs in systems that are unstable once the breaking process starts. Let us therefore investigate a common threshold distribution where the systems are stable to begin with: the uniform distribution on $[0,1)$.

\begin{figure}
\begin{center}
\includegraphics[width=\columnwidth]{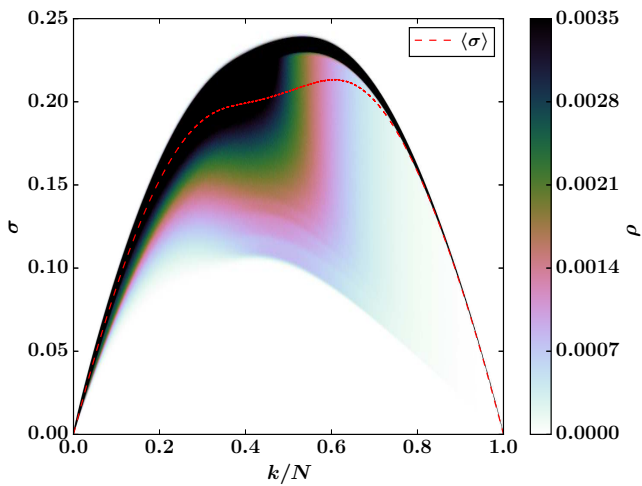}
\caption{The dashed, red line is the sample averaged load curve for the LLS model with a uniform threshold distribution on $[0,1)$, equation (\ref{eq:uniform}) with $t_{0} = 0$. It is based on simulations on a square lattice ($N = 128^{2}$). The background is a color map that shows the density $\rho$ of single sample LLS load curves for the $1.5 \times 10^{5}$ samples that the average is based on. The color bar is capped at $\rho = 0.0035$ to highlight the fluctuations with the smallest values of $\sigma$.}
\label{fig8}
\end{center}
\end{figure}

Fig.\ \ref{fig8} shows the density $\rho$ of force fluctuations and the corresponding sample averaged load curve for this uniform threshold distribution with LLS on a square lattice. Again the lower end of the fluctuations disappear in a region around the percolation threshold, which makes the sample average increase. Due to this effect, the averaged load curve has its maximum at $k/N \simeq 0.607$, whereas the maxima of individual load curves are distributed around a median damage $k/N \simeq 0.533$. The difference between these two maxima is clearly seen in Fig.\ \ref{fig8}.

In the intermediate region, the sample averaged load curve indicates stability --- via its positive slope --- when the fiber bundles are actually unstable. Hence, it cannot be trusted as an indicator of global stability. In general, stability --- both local and global --- is a property of individual samples that cannot be inferred from sample averages.

\subsection{The Shielding Effect}

Note that the ELS model becomes unstable at $k/N = 1/2$ for the uniform threshold distribution on $[0,1)$, which means that the LLS model, surprisingly, collapses later than the ELS model. This is due to a shielding effect that also has its origins in the geometry of the underlying lattice, but is otherwise unrelated to the statistical effect we have presented here. We discuss this at length in Part II \cite{kbph19}.

\section{Implications for Other Models}

The ELS and LLS models are the two extremes of load sharing, and other models, like the $\gamma$-model \cite{hmkh02} or the soft clamp fiber bundle model \cite{bhs02}, should exhibit behavior and phenomena somewhere between ELS and LLS. Intermediate load sharing rules can have infinite interaction ranges, but they should have finite \textit{effective} ranges of interaction. The longer this interaction range, the more the model resembles ELS, and conversely, the shorter it is, the more the model resembles LLS.

With an effective range of interaction significantly smaller than the system size, a model is expected to contain the apparent stability and its underlying cause. Instead of a narrow perimeter where fibers break, there will be a \textit{boundary layer} where fibers break, with width equal to the effective interaction range. Our argument for the disappearance of the lower end of the force fluctuations remains the same for such a model, except that it no longer happens around the percolation threshold. Instead, this effect occurs when the boundary layer permeates most of the lattice, and cannot expand into new areas to find weak fibers.

Note that for intermediate effective interaction ranges, this effect may be less pronounced than in the LLS model, but it should still be present. Therefore, the apparent stability presented here and its explanation should be considered a general feature of brittle fracture processes in disordered materials.

\section{Conclusion}

We have demonstrated a general mechanism resulting in the average force not being a reliable indicator of stability during fracture processes with local stress enhancement due to bias in the fluctuations around the average. We find that for several threshold distributions in the fiber bundle model, this mechanism gives an apparent stability, the illusion of stability due to an increasing average force even though individual systems are not stable. This apparent stability occurs around the site percolation threshold of the lattice for the systems we have studied in two to four dimensions.

\section*{Acknowledgements}
We thank Santanu Sinha for interesting discussions. We would also like to thank the reviewers for suggesting the use of the geometric mean with reference to Anderson localization. This work was partly supported by the Research Council of Norway through project number 250158 and its Centers of Excellence funding scheme, project number 262644. MH thanks the Swiss National Science Foundation for an early postdoc mobility grant, number 171982.

{}
\end{document}